\documentstyle[prd,aps,epsfig]{revtex}
\newcommand{\al}{\alpha}

\newcommand{\de}{\delta}

\newcommand{\ra}{\rightarrow}
\newcommand{\SS}{\mbox{$\cal S$}}
\newcommand{\GG}{\mbox{$\cal G$}}

\newcommand{\be}{\begin{equation}}
\newcommand{\ee}{\end{equation}}

\newcommand{\lsim}{\lesssim}
\newcommand{\bea}{\begin{eqnarray}}
\newcommand{\eea}{\end{eqnarray}}
\newcommand{\bean}{\begin{eqnarray*}}
\newcommand{\eean}{\end{eqnarray*}}

\newcommand{\bk}{{\bf k}}
\newcommand{\bx}{{\bf x}}

\newcommand{\bq}{{\bf q}}
\begin{document}
\draft
\preprint{\ 
\begin{tabular}{rr}
UGVA-DPT 1999/11-zzzz &  \\ 
astro-ph/0005087 & 
\end{tabular}
} 
\twocolumn
[\hsize\textwidth\columnwidth\hsize\csname@twocolumnfalse\endcsname 
\title{Skewness as a probe of non-Gaussian  initial conditions}
\author{Ruth Durrer$^1$, Roman Juszkiewicz$^{1,2}$ 
Martin Kunz$^{1,3}$ and Jean--Philippe Uzan$^{1,4}$}
\address{(1) D\'epartement de Physique Th\'eorique, Universit\'e de
Gen\`eve, 24 quai E. Ansermet, CH--1211 Geneva 4 (Switzerland).\\
(2) Copernicus Center, Bartycka 18, 00 716 Warsaw (Poland).\\
(3) Astrophysics Department, Oxford university, Keble road,
Oxford OX1 3RH (England).\\
(4) Laboratoire de Physique Th\'eorique, CNRS-UMR 8627, 
B\^at. 210, Universit\'e
Paris XI, F--91405 Orsay Cedex (France).}

\date{\today}
\maketitle

\begin{abstract}
We compute the skewness of the matter distribution arising from 
non--linear evolution and from non--Gaussian initial perturbations.
We apply our result to a very generic class
of models with non--Gaussian initial conditions  and we estimate  
analytically the ratio between the skewness due to non-linear clustering and
the part due to the intrinsic non-Gaussianity of the models.
We finally extend our estimates to higher moments.
\end{abstract}
]
{
The source of the initial density
fluctuations which have led to the formation of 
structure, observed in the Universe today is 
unknown. Determining its nature
will certainly be of utmost importance for the
fruitful relation between high energy physics and cosmology. 

In models which presently attract most attention, 
initial density fluctuations are generated 
during an inflationary phase. In the simplest
inflationary models, the initial fluctuations
obey Gaussian statistics. If this picture is
correct, the deviations from Gaussianity we observe
today were induced by nonlinear gravitational instability
~\cite{peebles73,roman}. However, it is also conceivable that the present
deviations from Gaussianity have two components:
gravitationally induced and intrinsic, coming
from the initial conditions rather than nonlinear dynamics
~\cite{silk,fry94,GaMa,pablo}.
Here, we investigate to what extent an intrinsic component can be 
`washed out' by nonlinear
dynamics and on which scales it could be either detected
or constrained from above in future galaxy surveys.

We start by deriving a general expression
for the so-called skewness parameter, $S_3$, 
including the effect of an initial non--Gaussianity,
non--linear evolution and smoothing. We then estimate the
normalized $N$--point cumulant, $S_N$, for a wide class of models
and compare it with the result obtained in  Gaussian models due 
to mild non--linearities. 

If the galaxies trace the spatial mass distribution,
galaxy surveys ~\cite{review} can be used to estimate 
the cumulants of the mass density contrast field, given by 
\be\label{npoint}
M_N(R)\equiv\left\langle (\de_R)^N(\bx,\eta_0)\right\rangle_c
\ee
of the smoothed density field 
$\de_R(\bx,\eta)\equiv\int d^3\bx' W_R(|\bx-\bx'|)\de(\bx',\eta)$,
where $\delta(\bx,\eta)$ is the density field,  $\eta$ and $\eta_0$
the conformal time and its value today, and $W_R$ is a window function
(e.g. Gaussian or top--hat) of width $R$.
The brackets in (\ref{npoint})
denote an ensemble average and the subscript $c$  indicates
that we deal with the connected part of the $N$--point function. For a
Gaussian field, all cumulants of order $N > 2$ vanish:
$M_N=0$.  $M_2$ is the variance while $M_3$ is a measure of
the asymmetry of the distribution, known as skewness.
We will also use the more common normalized cumulant,
\[ S_N(R) =M_N(R)/(M_2(R))^{(N -1)} ~.\]
This ratio is constant (independent of $R$) in the weakly non-linear
regime~\cite{fry84}. 
To calculate the general expression
for $M_3(R)$ in the weakly nonlinear regime, we follow the method
developed in \cite{roman}. Expanding $\delta({\bf x},\eta)$ in
a perturbative series, $\delta_1+\delta_2
+{\cal O}(3)$ and solving the system of coupled
Euler, Poisson and continuity
equations at second order leads, in Fourier space, to
$\Delta_1(\eta,{\bf k})={ D}(\eta,{\bf k})$ and
$$
{\textstyle
\Delta_2 (\eta,{\bf k}) = (2\pi)^{-3/2}\,\int d^3{\bf q}
J({\bf q},{\bf k}-{\bf q})D(\eta,{\bf q})
D(\eta,\bf k-\bf q)
}
$$
where we consider only the fastest growing modes
and we use the convention 
$$
\textstyle{
\Delta_N(\eta,{\bf k}) =
(2\pi)^{-{3/2}}\,\int \delta_N(\eta,{\bf x}){\rm e}^{-i{\bf k \cdot x}}
d^3{\bf x} \; .
}
$$
At late times where
a possible source term or seed has decayed, the time and space dependence
of the function $ D$ can be factorized,  
$D(\eta,{\bf k}) = D_{+}(\eta)\varepsilon({\bf k})$, where $D_{+}$
is the standard linear growing mode ~\cite{peebles73}. Perturbation theory
gives \cite{roman}
\be\label{J}
{\textstyle
J({\bf k,q}) = \frac{2}{3}(1+\kappa)
  +(q/k)P_1(\mu)+\frac{2}{3}\left(\frac{1}{2}-\kappa\right)P_2(\mu),
}
\ee
where the $P_\ell$ is the Legendre polynomial of order $\ell$,
$\mu\equiv {\bf k}\cdot{\bf q}/kq$. The quantity
$\kappa$ is a weak function of $\Omega$; for $\Omega > 0.01$,
$\kappa \approx (3/14)\Omega^{-0.03}$ ~\cite{francois}. 
The smoothing applies order by order. In Fourier space, we have
$\Delta_{R}(\eta,{\bf k}) = D(\eta,{\bf k})W_k, $
$W_k$ being the Fourier transform of the window function.
To fifth order, the skewness is 
\begin{eqnarray}\label{S2S3}
M_3&=&\left<\delta_{R,1}^3\right>+
        3\left<\delta_{R,1}^2\delta_{R,2}\right>+{\cal O}(5).
\end{eqnarray}
We introduce the two--,three-- and four--point power spectra as
$\left<12\right>\equiv
{\cal P}_2(k_1)\delta(\bk_1+\bk_2)$,
\begin{eqnarray}
&\left<123
\right>\equiv
{\cal P}_3(\bk_1,\bk_2)\delta(\bk_1+\bk_2+\bk_3) ,\nonumber\\
&\left<1234\right>_c \equiv 
{\cal P}_4(\bk_1,\bk_2,\bk_3)\delta(\bk_1+\bk_2+\bk_3+\bk_4).
\label{power}
\end{eqnarray}
(The Dirac $\de$ is a simple consequence of statistical homogeneity which we
assume throughout.) 
Here $\left< 12\ldots N\right> \equiv \left< D(\eta,\bk_1)
D(\eta,\bk_2) \ldots  D(\eta,\bk_N) \right>$.
The functions ${\cal P}_2$ and ${\cal P}_3$ are also known as
the power spectrum and the bispectrum, respectively.
Inserting the Fourier transforms of $\delta_1$ and
$\delta_2$  after smoothing  in
(\ref{S2S3}), expressing the correlators of $D$ in terms of the
power spectra (\ref{power}) and performing one integration using
the Dirac function in (\ref{power}), we obtain
\bea \label{zetafin}
\lefteqn{
M_3(R) =  \int\frac{d^3\bk d^3\bq}{(2\pi)^6}{\cal P}_3(\bk,\bq)
	W_kW_qW_{|\bk+\bq|}
} \nonumber \\  &&
+\int\frac{d^3{\bf k}d^3{\bf q}}{(2\pi)^6}{\cal P}_2(k){\cal P}_2(q)
W_{k}W_{q}W_{|{\bf k}+{\bf q}|}J({\bf k},{\bf q}) + 
 \nonumber \\  &&
\int\frac{d^3{\bf k}d^3{\bf q}d^3{\bf
p}}{(2\pi)^6}{\cal P}_4({\bf k},{\bf q}\!-\! {\bf k},{\bf
p})W_{q}W_{p}W_{|{\bf q}+{\bf p}|}
J({\bf k},{\bf q} \! -\! {\bf k})
\eea
For a Gaussian field,  ${\cal P}_4 = {\cal P}_3 = 0$
and  the only non-vanishing contribution comes from the second
term in the above expression. 
For a top hat window, this term gives
$M_3 = (34/7 - \gamma)M_2^2$, with $\gamma =
- d\log M_2(R)/d\log R$ ~\cite{roman}. Note
also that $\gamma(R)$ is the logarithmic slope 
of the two-point correlation function of the
density fluctuations - the Fourier transform
of ${\cal P}_2(k)$. It is usually assumed that
$\gamma > 0$ (condition of hierarchical clustering,
see e.g. ~\cite{peebles73}).

The class of models we want to analyze are those where
fluctuations in the dark matter are induced by the energy and momentum 
of an  inhomogeneously distributed component which
contributes only a small fraction to the total energy momentum tensor
and which interacts only gravitationally with the cosmic fluid. Such a
component is denoted as `seed' \cite{durrer94}. As stressed above,
we need to compute the $N$--point power spectra of the density field
at the end of the linear regime. The comoving linear density fluctuation $D$ 
of the cosmic matter--radiation fluid  evolves according to
\cite{kunz99,durrer94}
\bea\label{s1} 
&&\ddot { D}+H\left(1-6w+3c^2_s\right)\dot { D} +
k^2 c^2_s D \nonumber \\
&&\quad\quad -
{\textstyle \frac{3}{2}}\left(1+8w-3w^2
-6c^2_s\right)H^2D = \SS(\bk,\eta),
\eea
with $\SS\equiv(1+w)4\pi G(f_\rho+3f_P)$,
$f_\rho$ and $f_P$ being the inhomogeneous energy density and
pressure of  the seeds. When the seed is a scalar field $\phi$ with
vanishing potential,
 $f_\rho+3f_P =\dot\phi^2$. $G$ is Newton's constant,
$a$ denotes the cosmic scale factor,
a dot refers to  the derivative with respect to conformal time, 
$H\equiv \dot a/a$, 
$w\equiv P/\rho$ and $c_s^2\equiv\dot P/\dot\rho$ are respectively the 
enthalpy and the adiabatic sound speed of the cosmic fluid.

Equation (\ref{s1}) can be solved by  a Green's function, $\GG$,
\be
D(\bk,\eta)= \int_{\eta_i}^\eta \GG(\bk,\eta,\eta') \SS(\bk,\eta')d\eta' ,
\label{sol1}
\ee
where $\eta_i$ is some early initial time deep in the radiation era.
For the linear part of the reduced $N$--point function we then obtain
\bea
&&\langle { D}(\bk_1,\eta)\cdots  { D}(\bk_N,\eta)\rangle_c = 
\int_{\eta_i}^\eta d\eta_1\cdots d\eta_N
\nonumber\\
&& {\scriptsize \GG(\bk_1,\eta,\eta_1)\cdots \GG(\bk_N,\eta,\eta_N) 
\langle \SS(\bk_1,\eta_1)\cdots \SS(\bk_N,\eta_N)  \rangle_c}.
\label{Npoint}
\eea
We define the connected $N$--point function of the source by
\bean
 \langle\SS(1)\cdots \SS(N)\rangle_c\equiv
 F_N(\bk_1,\cdots\bk_N;\eta_1\cdots \eta_N)\de\left(\sum\bk_i\right),
\eean
where $(i)\equiv({\bf k}_i,\eta_i)$.
Again, the $\de$ function of the sum of all momenta is a consequence of the
statistical homogeneity. 

 We now assume that the reduced $N$-point
function of the source can be replaced  by its `perfectly
coherent approximation' given by
\bea
&&F_N(\bk_1,\dots,\bk_{N-1};\eta_1,\dots, \eta_N) \simeq 
	{\rm sign}(F_N)\times	\nonumber\\ 
&&{\scriptsize\sqrt[N]{|F_N(\bk_1,\dots,\bk_{N-1};\eta_1,\dots, \eta_1)\cdots
F_N(\bk_1,\dots,\bk_{N-1};\eta_N,\dots, \eta_N)|}}
\label{cohe}
\eea
(here and below, $\bk_N$ is always given by $\bk_N=-(\bk_1+\cdots
+\bk_{N-1})$). 
This approximation is exact if the evolution equation for $\SS$ is
linear and the randomness is entirely due to initial conditions.
Then the source term is of the form $\SS(\bk,\eta)=
R(\bk)s(k,\eta)$, where only $R$ is a random variable and $s$ is a
deterministic solution to the linear evolution equation of $\SS$ which
can be taken out of the average $\langle\rangle$. This is the key
property which renders the $N$-point function decoherent. Then
$F_N$ can be written as
\bea
&&\lefteqn{F_N(\bk_1,\dots,\bk_{N-1};\eta_1,\dots, \eta_N) \simeq }\nonumber
\\ 
&& s(1)\cdots s(N)\langle R(\bk_1) \cdots R(\bk_N)\rangle_c
\eea
which is clearly of the form (\ref{cohe}).

An important example are models with no sources but with
non-Gaussian initial conditions for $D$. Such models, like {\em e.g.} 
the recent $\chi^2$ Peebles model~\cite{peebles99}, are always
perfectly coherent and therefore included in our analysis: In this case
$D(\bk,\eta)=R(\bk)d(k,\eta)$, where $R$ is a non-Gaussian random
variable given by the initial condition and $d$ is a deterministic 
homogeneous solution of Eq.~(\ref{s1}). Clearly, if we choose 
$\SS(\bk,\eta) = R(\bk)\de(\eta-\eta_{in})$ and $\GG(k,\eta,\eta')=
d(k,\eta)$,  $D$ is of the form (\ref{sol1}). Therefore, models where
the non-Gaussianity is purely due to initial conditions are always
perfectly coherent. As the equation of motion for $D$ is second order,
the homogeneous solution has in principle two modes,
$D=R_1(\bk)d_1(k,\eta) +R_2(\bk)d_2(k,\eta)$, but since we shall
evaluate the $N$-point functions deeply in the matter era, the
decaying mode will have disappeared and may thus be neglected in our analysis. 

Models where the source term is due to a scalar field which evolves
linearly in time are not perfectly coherent, since $\SS$ is given by the
components of the energy momentum tensor which are quadratic in the
fields. Numerical calculations, however, have shown that this
non-linearity is not severe and perfect coherence is a relatively good
approximation~\cite{DKM99,kunz97}. One
example of this kind are axionic seeds in pre-big bang
cosmology~\cite{durrer98,melch99,vern00} for which decoherence has
been tested and is found to be on the level of less than 5\% for the CMB
power spectrum.
In Fig.~1 the  functions $D_2(k,\eta)$ and $D_3(k,k,\eta)$ as
obtained by a full numerical calculation are compared to  their coherent
approximation (\ref{cohe}) for the large-N limit of  global $O(N)$
symmetric scalar fields. This is another example where the scalar field
evolution is linear and the only non-linearity in the source term is
due to the energy momentum tensor being quadratic in the
field~\cite{TS,kunz97,DKM99}. 
\begin{figure}[ht]
\centerline{\epsfig{file=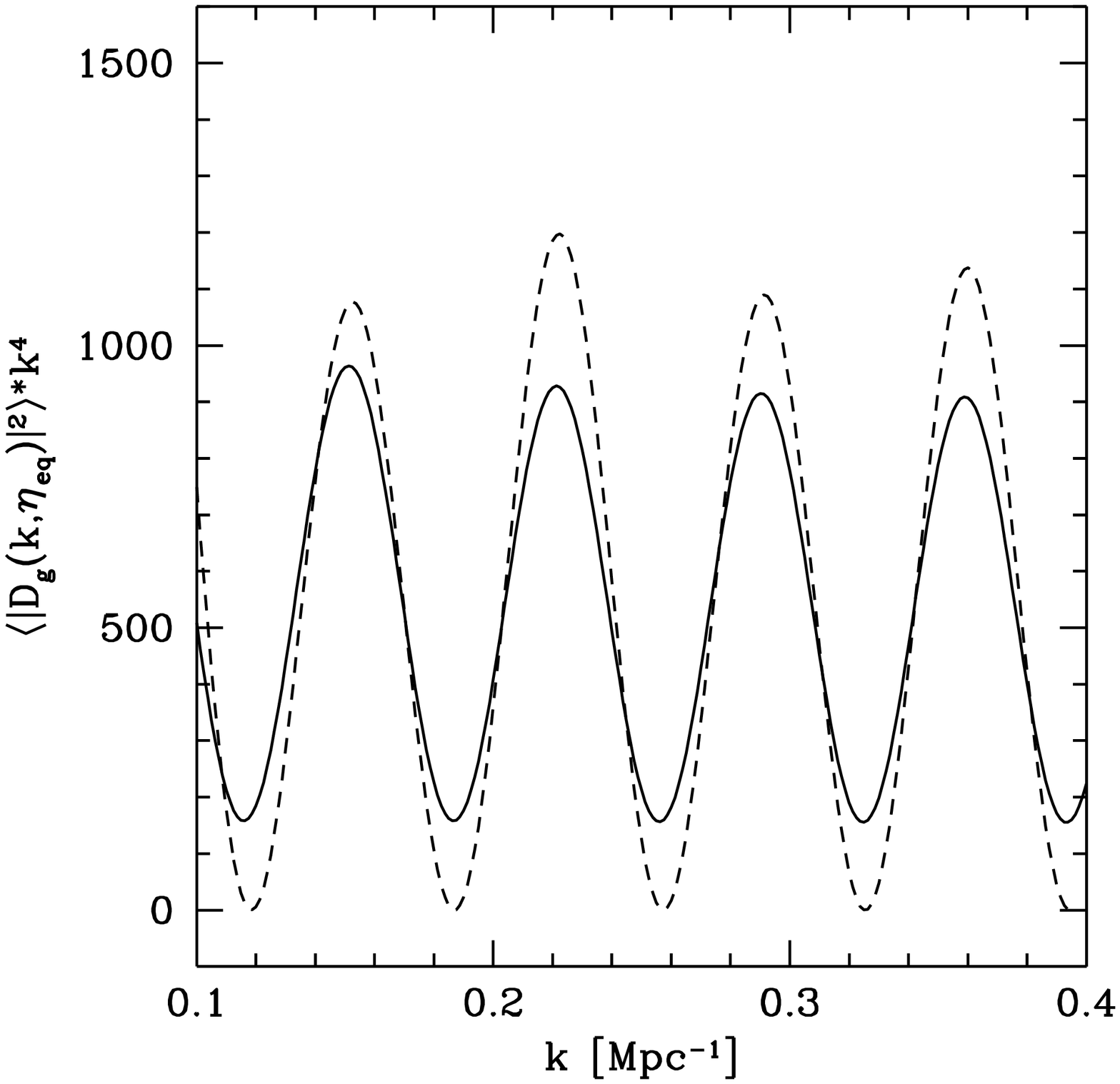, width=2.5in}}
\centerline{\epsfig{file=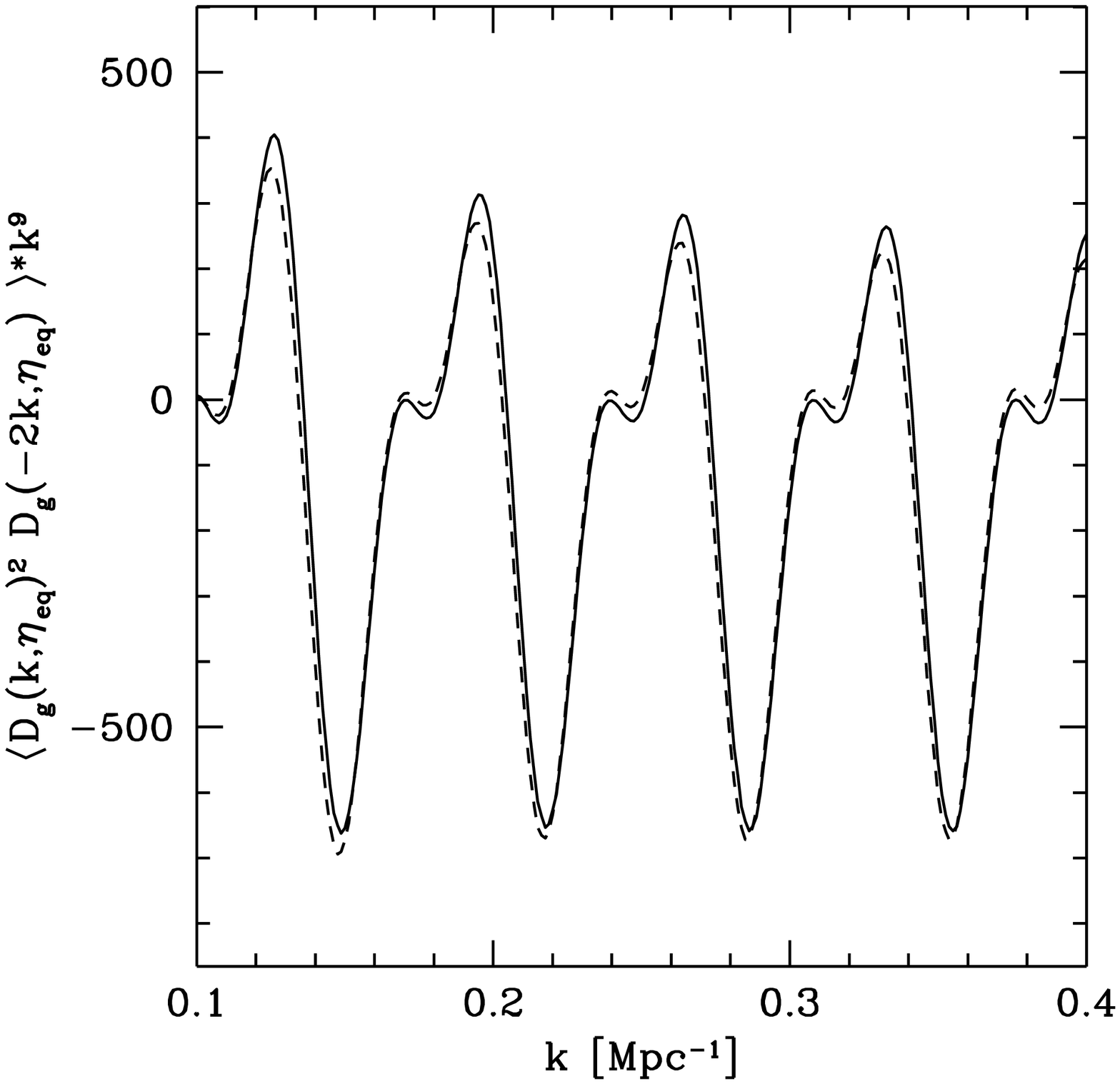, width=2.5in} }
\caption{The coherent approximation (dashed line) and the full
decoherent result (solid line) for the 2- (top) and 3-point (bottom)
functions of the large-N limit of global $O(N)$ symmetric scalar
fields is shown at the end of the radiation era. The sign in the
coherent approximation for the 3-point function is chosen to agree
with the sign for the decoherent 3-point function.}
\end{figure}
 
For topological defects, especially for cosmic strings, the perfectly coherent
approximation misses several important features (like the `smearing
out' of secondary acoustic peaks). However, we believe that our
generic scaling result holds also in this case, as is indicated by
numerical simulations of global texture:
even though global texture show considerable
decoherence~\cite{DKM99}, the same scaling law for higher moments which we
derive here has been discovered numerically~\cite{GaMa}. 

Under the perfectly coherent approximation Eq.~(\ref{Npoint}) can be
factorized as the product of the $N$ solutions, 
$D_{Nj}(\bk_1, \dots, \bk_{N-1},\eta)$ of the equations (\ref{s1}) with
source term $[F_N(\bk_1,\dots, \bk_{N-1}; \eta,\dots, \eta)]^{1/N}$, where
$\bk_j$ is the wave number $\bk$ appearing in the term $c_s^2k^2$ 
on the left hand side of
(\ref{s1}) and the other wave numbers have to be considered like
parameters of the source term,
\bea
&&\langle D(\bk_1,\eta)\cdots  D(\bk_{N},\eta)\rangle_c \simeq  \nonumber \\
&& \left[\prod_{j=1}^ND_{Nj}(\bk_1, \dots, \bk_{N-1},\eta)
\right]\de(\sum \bk_i) \nonumber \\
&& \qquad\quad\equiv {\cal P}_N(\bk_1,\dots,\bk_{N-1},\eta)
\de(\sum\bk_i).
\eea

To continue, we assume that $F_N$ is a simple power law in the $k_i$ on
super-Hubble scales and that it decays after Hubble crossing. This
behavior is certainly correct for all examples discussed in the
literature so far. We can then make the following ansatz 
\be
    F_N \simeq\left\{\begin{array}{ll}   
     \prod_{n=1}^N{k_n^\al \over k_0^\al} (f(\eta)\eta)^N\eta^{-3} & 
	\mbox{if } k_i\eta\le 1, \forall  i\in \{1,\dots,N\} \\
       0 & \mbox{otherwise}~.
\end{array}
\right.
\ee 
Here $f$ is a dimensionless function and $k_0$ is an arbitrary scale. 
For scale invariant
seeds (e.g. topological defects) $f$ is just a constant and $\al=0$.
For axion seeds generated during a pre-big bang phase, $\al$ depends
on the spectral index of  the axion field, which in turn is determined
by the evolution law of the extra dimension~\cite{melch99}. For the
Peebles model $\al$ is given by the power spectrum of the scalar field
$\phi$ and $f$ is a delta-function.
Since $F_N$ is symmetrical in the variables $\bk_j$ we can order
them such that $k_1\ge k_2\ge \cdots \ge k_{N}$.

Let us discuss the temporal  behavior of the variables $D_{Nj}$. As long as
$k_1\eta < 1$, the term $c_s^2k_j^2D$ can be neglected in
Eq.~(\ref{s1}) and the Green's function is a power law.
At $k_1\eta \sim 1$ the source term decays and as long as a
perturbation remains super horizon, it just grows like $ \eta^2$, so
that  for  $k_j\eta<1<k_1\eta$,
\[
 D_{Nj} \approx g(1/k_1)k_1^{-2+3/N}(\eta k_1)^2 \Pi_{n=1}^N(k_n/k_0)^\al 
\]
where
\[ g(\eta)= {4\pi G\over \eta^{2-3/N}}\int^\eta_{\eta_{in}}{
	G}(\eta,\eta')f(\eta')\eta'^{(2-3/N)}{d\eta'\over \eta'} ~,
\]
and we have to take the part of the integral above which converges
when $\eta_{in}\ra 0$.

Once the perturbation enters the horizon it either starts oscillating
with roughly constant amplitude or continues to grow $\propto \eta^2$,
depending on whether $k_j$ enters during the radiation or matter
dominated era. At late time, $\eta\gg\eta_{eq}$ and $k\eta\gg1$, we
therefore obtain
\bean  
D_{Nj} &\approx&  g(1/k_1)k_1^{-2+3/N}(k_1/k_j)^2 \\
&& \prod_{n=1}^N(k_n/k_0)^\al
    \left\{\begin{array}{ll} \left({\eta\over \eta_{eq}}\right)^2 &
    \mbox{if } k_j\eta_{eq}>1 \\
     (\eta k_j)^2 &  \mbox{if } k_j\eta_{eq}< 1 \\
   \end{array} \right.
\eean
where $\eta_{eq}$ is the time of equality between the
 matter and radiation densities.
Defining $0\le j_{eq}\le N$ so that $k_j\eta_{eq}>1$ for all $j\le
j_{eq}$ we obtain for the connected $N$-point function

\bea
{\cal P}_N(\bk_1,\dots,\bk_{N-1},\eta) &\simeq& g(1/k_1)^Nk_1^3\eta^{2N}
\nonumber\\
&&	 \prod_{n=1}^N\left({k_n\over k_0}\right)^\al\prod_{j=1}^{j_{eq}}
	\left({1\over k_j\eta_{eq}}\right)^2
\eea
Using this result for the ordinary power spectrum, ${\cal P}_2$, we can
express ${\cal P}_N$ is terms of products of ${\cal P}_2$ as
\bea
 &&{\cal P}_N(\bk_1,\cdots,\bk_{N-1},\eta) \simeq \nonumber \\
 &&  k_1^{3(1-N/2)}
\prod_{j=1}^N\left(\sqrt{{\cal P}_2(k_j,\eta)}{ g(1/k_1)k_j^{3/2}\over
	g(1/k_j)k_1^{3/2}}\right)
 . \label{PN}
\eea
For the class of models considered and under the
assumption of perfect coherence, we have determined the connected
$N$--point power spectra in the linear regime which are the input
of the skewness (\ref{zetafin}).

$M_3$ has two contributions: A linear one due to the initial
non--Gaussianity (contained in ${\cal P}_3$) and one due to
non--linear clustering which induces skewness even in an originally
Gaussian distribution of perturbations; it contains a Gaussian part
(${\cal P}_2^2$) and a non-Gaussian term (${\cal P}_4$). We decompose 
the skewness as 
$$M_3=M_3^{(L)} + M_3^{(NL)}$$
We want to estimate the ratio of these two contributions. Under
our approximation~(\ref{PN}), the first term of (\ref{zetafin})
reduces to
\bea
&&M_3^{(L)}=\int\frac{d^3{\bf k}d^3{\bf k}'}{(2\pi)^6}W_kW_{q}
  W_{|{\bf k}+{\bf k}'|}\sqrt{{\cal P}_2(k){\cal P}_2(q){\cal P}_2(|\bk+\bq|)}\nonumber \\
&&\qquad\qquad k_{\max}^{-3/2}\left[ g(1/k_{\max})^3
 (kq|\bk+\bq|)^{3/2}\over
   g(1/k) g(1/q)g(1/|\bk+\bq|)k_{\max}^{9/2}\right] ,
	\label{S3lin}
\eea
where $k_{\max}\equiv\max\{k,q,|\bk+\bq|\}$. 
$M_3^{(NL)}$ is given by the second and third terms in (\ref{zetafin}).

To estimate analytically the ratio 
$M_3^{(L)}/M_3^{(NL)} = S_3^{(L)}/S_3^{(NL)}$,
we make the following approximations:
\begin{itemize}
\item[-] We assume that  ${\cal P}_2$  is a simple power law within the
range of scales  of interest, namely all the modes which enter the
horizon during the radiation era, this is
   $0.1h^{-1}$Mpc$ \lesssim 2\pi/k \lesssim 20h^{-2}$Mpc, name. 
${\cal P}_2(k) = k^{-3}(k/k_*)^{\gamma} ~.$
\item[-] We also assume that $g(\eta) \propto \eta^r$.
\item[-] We replace the window function by 
a simple cut--off at $k=1/R$.
\item[-] For symmetry reasons we may integrate over 
the triangle $q\le k\le R$ and then multiply the result by 2.
\item[-] Since in our integration region, $q\le k$, we replace
$|\bk+\bq|$ by $k$.
\end{itemize}

With these approximations the angular dependence of the integrand
disappears and the integrals over $\bk$ and ${\bf q}$ in (\ref{zetafin}) can be
trivially performed leading to
\bea
 M_3^{(L)}(R) &\simeq&  {4 (k_*R)^{-3{\gamma}/2}\over (2\pi)^4 3{\gamma}(3+{\gamma}/2+r)} \\
 \mbox{ for }	&&   {\gamma}>0 ~ \mbox{ and }~3+{\gamma}/2+r>0  \nonumber \\
 M_3^{(NL)}(R) &\simeq& { (k_*R)^{-2{\gamma}}\over (2\pi)^4 {\gamma}^2} \quad
 \mbox{for}\quad {\gamma}>0, \nonumber
\eea
where we have just considered the Gaussian contribution, ${\cal
P}_2^2$ to $M_3^{(NL)}$.

Since $k_*$ is just the scale beyond which the density contrast
$\langle { D}(\bx)^2\rangle_{R=1/k} \sim {\cal P}_2(k)k^3$ is larger than unity
and non-linearities become important, we define the non--linearity
scale $R_{\rm lin}=1/k_*$. The ratio between the skewness due to the
non--Gaussianity in the linear perturbation and 
the one due to dynamical nonlinearities
is then 
\be
 { S_3^{(L)} \over  S_3^{(NL)}}\sim {4{\gamma}\over 3(3+{\gamma}/2-r)}  
 \left({R\over R_{\rm lin}}\right)^{{\gamma}/2}\label{rat3}~.
\ee
This is our  main result. 
 It is readily checked that the non-Gaussian contribution, ${\cal
P}_4$, to $M_3^{(NL)}$ behaves just like the contribution 
$M_3^{(NL)}$ and thus only modifies the pre-factor in
(\ref{rat3}), which should  not be taken too
seriously in view of the relatively crude approximations which we have
employed to obtain our result.

This computation of the skewness is easily generalized to higher
moments. As our computation shows, linear non--Gaussianities scale
like \be M_N^{(L)}(R) \propto (R/R_{\rm lin})^{-N{\gamma}/2} ~.  \ee
The dominant non--linear contribution to the {\bf connected} $N$-point
function which is also present in Gaussian theories contains $N-2$
second order terms ${ D}_2$~\cite{fry84} and therefore scales like \be
M_N^{(NL, {\rm Gauss})}(R) \propto (R/R_{\rm
lin})^{-(N-1){\gamma}}~. \label{NGauss} \ee The lowest order
non-linearity for a generic non-Gaussian model, however just comes
from the non-Gaussian term with $N+1$ factors of $D$. The non-Gaussian
non-linear corrections therefore generically scale like \be
M_N^{(NL,{\rm noGauss})}(R) \propto (R/R_{\rm lin})^{-(N+1)\gamma/2}~.
\label{NnoGauss} \ee Only for $N=3$ the two terms (\ref{NGauss}) and
(\ref{NnoGauss}) scale in the same way. For all higher $N$'s the
non-Gaussian contribution dominates in the mildly non-linear regime,
$R\ge R_{\rm lin}$.  From Eq.~(\ref{NnoGauss}) we infer that in on
large scales the ratios for all reduced $N$-point functions very
generically scale like \be {S_N^{(L)}(R)\over S_N^{(NL)}(R)} \propto
\left({R\over R_{\rm lin}}\right)^{{\gamma}/2}~. \label{ratN} \ee This
expression agrees with other analytic predictions~\cite{silk} as well
as numerical simulations in a global texture model~\cite{GaMa}. The
agreement with the texture simulations which are decoherent suggests
that the validity of our result extends beyond the conditions under
which Eq.~(\ref{ratN}) was derived.  More important than decoherence
is that the source term decays at late times and therefore the density
perturbations just evolve according to the homogeneous solution. This
implies that at late times the $N$-point functions behave like the
homogeneous growing mode to the $N$th power, while the reduced
$N$-point function induced by non-linear clustering from Gaussian
perturbations scales like the growing mode to the $2(N-1)$th
power. Since topological defect sources decay on sub-horizon scales,
we conclude that the derived scaling behavior is also valid for them
(this argument will be expanded in our follow up
publication~\cite{kunz99}).

Our result implies that on small scales ($R\lsim R_{\rm lin}$),
the dominant contribution to the cumulants comes from nonlinear
Newtonian gravitational clustering, and the Gaussian term actually
dominates. Intrinsic deviations from Gaussianity are
difficult to detect on small scales. Hence, we should look for 
signs of intrinsic non-Gaussianity at large
scales ($R > R_{\rm lin}$). This suggestion was expressed earlier
based  on qualitative physical arguments~\cite{silk}; however, our
present  result is derived from first principles for a specific
class of initial conditions -- coherent seeds.

If galaxies trace mass, the measurements of the two-point
correlation function suggest $R_{\rm lin}\sim 10h^{-1}$Mpc 
and $\gamma(R) \approx 1.8$ for $10 {\rm kpc} \lsim hR \lsim 15$ Mpc
(here $h$ is the usual parameterization for the Hubble constant
in units of 100 km s$^{-1}$Mpc$^{-1}$); the slope $\gamma$ becomes
steeper at larger separations $R$~\cite{peebles73,review}.
A frequently considered theoretical possibility for
long-wave tail of the initial ${\cal P}_2(k)$, called the
Zel'dovich-Harrison spectrum, would give $\gamma = 4$ at large
separations. Hence, we can expect all $S_N$s to ``blow up''
with increasing scale for the class of non-Gaussian
models considered here, in contrast with models with
Gaussian initial conditions. The available measurements of $S_3(R)$
and $S_4(R)$ do not show such a rise with scale and have
already been used to constrain texture models~\cite{GaMa}.
Likewise, there are indications that the existent
data from the APM Galaxy Survey may may already extend
to sufficiently large scales to constrain the $\chi^2$ Peebles model 
\cite{josh}. With surveys presently underway like
the Sloan Digital Sky Survey~\cite{sloan}, the prospects for
using the approach outlined here to probe the statistics
of the cosmological initial conditions will become even better.

In this work we derived a scaling law for the ``intrinsic
to induced'' skewness ratio (\ref{rat3})
for coherent seeds. We also showed how to generalize
this law to higher cumulants. We plan to follow these calculations
with more detailed predictions for coherent seed models and to
confront our analytic results with 
numerical simulations as well as observational data from 
galaxy surveys \cite{kunz99}. Let us also repeat that the derived
scaling laws seem to be more general than their derivation as they
have been obtained numerically for global texture which are decoherent
seeds. We actually believe that the origin of the scaling laws is not
coherence but mainly the decay of the sources at late time and we
therefore conjecture that they hold also for topological defects.

}

\begin{references}
\bibitem{peebles73}P.J.E. Peebles, {\it The Large Scale Structure of 
	the Universe}, Princeton University Press (1980).
\bibitem{roman} R. Juszkiewicz, F. Bouchet and S. Colombi,
	Astrophys. J. Letters {\bf 412} (1993) L9.
\bibitem{silk} J. Silk and R. Juszkiewicz, Nature {\bf 353} (1991) 386.
\bibitem{fry94} J.N. Fry and R.J. Scherrer, Astrophys. J. {\bf 429} (1994)
	36.
\bibitem{GaMa} E. Gazta{\~n}aga and P. M{\"a}h{\"o}nen,
	Astrophys. J. {\bf 462} (1996) L1.
\bibitem{pablo}  E. Gazta{\~n}aga and P. Fosalba, Mon. Not. 
	Roy. Astron. Soc. {\bf 301} (1998) 524.
\bibitem{fry84} J. Fry, Astrophys. J. {\bf 279}, 499 (1984).
\bibitem{francois} F. Bouchet, R. Juszkiewicz, S. Colombi and
	R. Pellat, Astrophys. J. {\bf 394} (1992) L5.
\bibitem{kunz99} M. Kunz, J-Ph. Uzan, R. Durrer and R. Juszkiewicz,
	in preparation (2000).
\bibitem{durrer94}  R. Durrer, Phys. Rev. D {\bf 42} (1990) 2533.
\bibitem{peebles99} P.J.E. Peebles, Ap. J. {\bf 510} (1999) 523;
	{\it ibid.}, Ap. J. {\bf 510} (1999) 531.
\bibitem{DKM99}  R. Durrer, M. Kunz and A. Melchiorri, Phys. Rev.
	{\bf D59}  (1999) 123005. 
\bibitem{kunz97}  M. Kunz and R. Durrer, Phys. Rev. {\bf D55} (1997) R4516.
\bibitem{durrer98} R. Durrer, M. Gasperini, M. Sakellariadou and
	G. Veneziano, Phys. Lett. {\bf B436} (1998) 66; {\it ibid.}, 
	Phys. Rev. {\bf D59} (1999) 043511.
\bibitem{melch99} A. Melchiorri, F. Vernizzi, R. Durrer and
	G. Veneziano, Phys. Rev. Lett. {\bf 83}, 4464 (1999) [{\tt
	astro-ph/9904167}]. 
\bibitem{vern00} F. Vernizzi, A. Melchiorri, R. Durrer and
	G. Veneziano, in preparation.
\bibitem{TS}N. Turok and D. Spergel, Phys. Rev. Lett. {\bf 66}, 3093
	(1991).
\bibitem{review} M.A. Strauss and J.A. Willick, Phys. Rep. {\bf 261}
	(1995) 271.
\bibitem{josh} J. Frieman and E. Gazta{\~n}aga, Astrophys. J. 
	{\bf 521} (1999) L83.
\bibitem{sloan}J. Loveday and J. Pier, Proceedings of the
	14th IAP meeting {\em Wide
     	field surveys in cosmology}, ed. Y. Mellier et al. (1998)  [{\tt
	astro-ph/9809179}]. 
\end{references}
\end{document}